\documentclass[twocolumn,showpacs,preprintnumbers,amsmath,amssymb,prb]{revtex4}

\usepackage{graphicx}
\usepackage{dcolumn}
\usepackage{bm}

\begin{document}

\title{Time-dependent density-functional theory of exciton-exciton
correlations in the nonlinear optical response}
\author{Volodymyr~Turkowski},
\author{Michael~N.~Leuenberger}
\altaffiliation{Corresponding author, e-mail address: mleuenbe@mail.ucf.edu}

\affiliation{Department of Physics and NanoScience and Technology Center,
University of Central Florida, Orlando, FL 32816}

\date{\today}

\begin{abstract}
We analyze possible nonlinear exciton-exciton correlation effects in the optical response
of semiconductors by using a time-dependent density-functional theory (TDDFT) approach. 
For this purpose, we derive the nonlinear (third-order) TDDFT equation for the excitonic 
polarization. In this equation, the nonlinear time-dependent effects are described by the time-dependent (non-adiabatic) part of the 
effective exciton-exciton interaction, which depends on the exchange-correlation (XC) kernel. 
We apply the approach to study the nonlinear optical response of a GaAs quantum well. 
In particular, we calculate the 2D Fourier spectra of the system and compare it with experimental data.
We find that the memory effects play a crucial role in this response, and in particular that it is necessary to use a non-adiabatic XC kernel to describe excitonic bound states - biexcitons, which are formed due to the retarded TDDFT exciton-exciton interaction.

\end{abstract}



\maketitle

\section{Introduction.}
\label{Introduction}

The nonlinear optical response of semiconductors
 is an important problem of modern condensed matter physics, 
in particular due to the necessity to describe correctly the ultrafast laser pulse experiments \cite{1},
including the four-wave mixing (FWM) case \cite{3}, in nanostructures and other systems. 
This problem is a part of a more general problem of the nonequilibrium nonlinear behavior of quantum matter
at ultrashort time scales, one of the most complicated and important problems of modern physics. It is extremely important
to understand this behavior from both fundamental science and
technological points of view, and the FWM experiments is one of the most powerful and promising tools
currently used for this purpose.
A fundamental process in the FWM experiments is generation of multiple excitonic states,
which can have variety of practical applications
from optoelectronic devices \cite{5} to quantum computing \cite{7}. 
In many cases, including the ultrafast response, the nonlinear effects, which come from the time-dependent exciton-exciton
interaction, are very important.
To our knowledge, these were studied only in the framework of many-body effective models by 
deriving and solving the effective third-order equation for the polarization
 (see e.g. Refs.~\cite{3,Ostreich,Ostreich3}). In particular, it was shown that the nonlinear effects come from
the time-dependent exciton-exciton interactions defined by the corresponding correlation functions.
Due to the complexity of the problem, the time-dependent correlation function is usually  approximated by a phenomenological expression. 
Since the optical response strongly depends on the form of this expression \cite{Ostreich3},
it is highly desirable to obtain the form of the effective interaction from a simple but fundamental approach that
does not include (or at least includes a minimal number of) phenomenological parameters. For this purpose,
we propose to use TDDFT \cite{12}.
It was already successfully applied to study the excitonic effects in the frameworks
of the time-dependent optimized effective potential approach \cite{14}, in combination with the
Bethe-Salpeter equation \cite{1}, and by solving the TDDFT version of the semiconductor Bloch equations (SBEs, see e.g. Ref.~\cite{8}) \cite{15,16}. We recently generalized the last approach to the biexcitonic case \cite{biexcitons}.
In most of these papers, the contribution of the exciton-exciton interaction
to the optical absorption spectrum was neglected. Only in Ref.~\cite{15} the nonlinear excitonic
effects were taken into account by solving a nonlinear system of the TDDFT-SBEs. However, it was difficult to
identify the nonlinear contribution from such a general solution. In addition, since the adiabatic approximation
was used, the time-resolved exciton-exciton interaction was neglected in this case.
As we show below, the non-adiabaticity of the XC kernels is a crucial requirement for the development of a theory based on TDDFT to
describe nonlinear correlation effects including the formation of exciton-exciton bound states (biexcitons).
Another requirement, which follows from all the studies mentioned above, is the
necessity to go beyond the LDA and GGA approximation in order to obtain even single-exciton effects. 

Currently, possible structures of the non-adiabatic XC potentials are much less understood comparing to the cases of static
DFT and adiabatic TDDFT. However, the frequency dependence of the XC potential is not only important for the description of the exciton-exciton interaction but also for various static problems such as multiple excitations \cite{biexcitons,17,18,19,24} and time-dependent
problems when the system response is analyzed (see, e.g., Ref.~\cite{Li}). In this work we use the available exact knowledge of the asymptotic limits
of the XC kernels at low and high frequencies and some experimental data to construct the non-adiabatic part of the kernels that are able to describe the effects
of the exciton-exciton correlations within TDDFT, including biexcitonic bound states. We test the approach by calculating the response of a  GaAs quantum well in a FWM experiment and demonstrate that the approach is capable to describe the main features of the spectrum.
The method proposed here can be used in a more general case to study the non-adiabatic response within TDDFT.

\section{Method}
\label{Method}

In order to derive the nonlinear equation for the polarization, we use the density-matrix 
representation of TDDFT. For simplicity, we consider the three-band case (two heavy- and light-hole valence bands and one conduction band), though the generalization to more bands
is straightforward. In the density-matrix TDDFT, one looks for the solution of the time-dependent Kohn-Sham (KS)
equation
\begin{eqnarray}
i\partial \Psi ({\bf r},t)/\partial t={\hat H}({\bf r},t)\Psi ({\bf r},t)
\label{KS}
\end{eqnarray}
 as a linear combination of the wave functions which corresponds to the static solution,
\begin{eqnarray}
E_{l}\psi^{l(0)} ({\bf r})={\hat H}({\bf r},t=0)\psi^{l(0)}({\bf r},t),
\label{projection}
\end{eqnarray}
 where $l$ is the band number.
In the last equation,
\begin{equation} 
{\hat H}({\bf r},t)=-\frac{\nabla^{2}}{2m}+V({\bf r},t)
+
V_{\rm H}[n]({\bf r},t) + V_{\rm xc}[n]({\bf r},t) 
\label{TDDFT_Hamiltonian}
\end{equation}
is the time-dependent KS Hamiltonian,
where we set $\hbar=1$, 
\begin{eqnarray}
V({\bf r},t) = V_{\rm nucl}({\bf r})+V_{\rm ext}({\bf r},t)
\label{V}
\end{eqnarray}
 is the sum of the static atomic potential
and the time-dependent external potential, and $V_{\rm H}({\bf r},t)$ and $V_{\rm xc}({\bf r},t)$
are the time-dependent Hartree and  XC potentials, which depend on the 
single-particle density $n({\bf r},t)$ (which should be found self-consistently
from the KS wave functions). We use the dipole approximation for the external perturbation arising from the electric field pulse,
\begin{eqnarray}
V_{\rm ext}({\bf r},t)=-{\bf r}E(t), 
\label{Vext}
\end{eqnarray}
which corresponds to the case of high frequencies, comparing to the spacing between the energy levels.
In the three-band case, the wave function can be approximated as a linear combination
of the valence and the conduction electron wave-functions:
\begin{eqnarray}
\Psi_{{\bf k}} ({\bf r},t)=\sum_{l=v1,v2,c}c_{{\bf k}}^{l}(t)\psi_{{\bf k}}^{l(0)}({\bf r}),
\label{Psi}
\end{eqnarray} 
and the problem reduces to the solution of the Liouville equation
\begin{equation}
i\frac{\partial \rho_{{\bf k}}^{lm}}{\partial t}=\left[H,\rho\right]_{{\bf k}}^{lm},
\label{Liouville}
\end{equation}
where
\begin{equation}
H_{{\bf k}}^{lm}(t) = \int d{\bf r}\psi_{{\bf k}}^{l(0)*}({\bf r})
{\hat H}({\bf r},t)
\psi_{{\bf k}}^{m(0)}({\bf r})
\label{Hlm}
\end{equation}
are the matrix elements of the Hamiltonian with respect to the static wave-functions
and 
\begin{eqnarray}
\rho_{{\bf k}}^{lm}(t)=c_{{\bf k}}^{l}(t)c_{{\bf k}}^{l*}(t)
\label{rho}
\end{eqnarray}
 is the $3\times 3$ density matrix, 
whose diagonal elements correspond to the band occupancies 
\begin{eqnarray}
n_{v_{i}}(t)&=&\rho_{{\bf k}}^{v_{i}v_{i}}(t), (i=l,h)
\label{nv}
\\
n_{c}(t)&=&\rho_{{\bf k}}^{cc}(t)=
1-\rho_{{\bf k}}^{v_{1}v_{1}}(t)-\rho_{{\bf k}}^{v_{2}v_{2}}(t),
\label{nc}
\end{eqnarray}
 and the non-diagonal elements - to the direct and the inverse 
(including excitonic) coherences $\rho_{{\bf k}}^{cv}(t)$ and  
$\rho_{{\bf k}}^{vc}(t)=\rho_{{\bf k}}^{cv*}(t)$. For simplicity, we take into account only the vertical transitions since in most cases the photon momentum can be neglected. The explicit form of the system of independent Liouville equations have the form of the SBEs \cite{8}
for the excited electron density $n_{c}(t)$ and polarization $\rho_{{\bf k}}^{cv}(t)$ (for details, see Ref.~\cite{14}).
We shall consider the nonlinear equation for the exciton polarization by taking into account the terms up to the third order
in $\rho_{{\bf k}}^{cv}(t)$ (or, more precisely, in $c_{{\bf k}}^{c}(t)$).
This can be done by expanding the nonlinear part $V_{XC{\bf k}}^{lm}(t)$ of the matrix elements 
$H_{{\bf k}}^{lm}(t)$ in powers of the fluctuating particle density 
\begin{eqnarray}
\delta n({\bf r},t)=n({\bf r},t)-n({\bf r},t=0),
\label{deltan}
 \end{eqnarray}
i.e.
\begin{eqnarray}
V_{XC{\bf k}}^{lm}(t)= \int d{\bf r}\psi_{{\bf k}}^{l(0)*}({\bf r})
\left(
\sum_{n}\frac{1}{n!}\frac{\delta^{n} V_{XC}}{\delta n^{n}}\delta n^{n}
\right)
\psi_{{\bf k}}^{m(0)}({\bf r}),
\nonumber \\
\label{VXC}
\end{eqnarray}
 where the expression in the brackets is the Taylor expansion of the functional $V_{XC}$, including the space-time integration
over the internal variables. As it follows from the definition of the wave-function and the density-matrix,
one can use the expansion of the wave function in terms of the $c$-coefficients in Eq. (\ref{Psi}) and express the density fluctuation in terms of the elements of the density matrix:
\begin{eqnarray}
\delta n({\bf r},t)&=&\sum_{k<k_{F}}\left(|\Psi_{{\bf k}}({\bf r},t)|^{2}-|\Psi_{{\bf k}}({\bf r},t=0)|^{2}\right)
~~~~~~~~~~~~~~~~
\nonumber \\
&\simeq&\sum_{k<k_{F}}\
[\rho_{{\bf k}}^{cv_{1}}(t)
\psi_{{\bf k}}^{c(0)}({\bf r})\psi_{{\bf k}}^{v_{1}(0)*}({\bf r})
\nonumber \\
&~&~~~~~~~~~+\rho_{{\bf k}}^{cv_{2}}(t)
\psi_{{\bf k}}^{c(0)}({\bf r})\psi_{{\bf k}}^{v_{2}(0)*}({\bf r})].
\nonumber \\
\label{deltan2}
\end{eqnarray}
In the last expression we neglect the complex conjugated terms,
since they correspond to the exciton de-excitations and are small
comparing to the presented terms when the photon frequency is close to the (usually large) energy gap.
\cite{15,16}
Next, taking into account the factthat the exciton (polarization) functions are related to the density matrix elements as
$P_{1{\bf k}}(t)=\rho_{{\bf k}}^{cv1}(t), P_{2{\bf k}}(t)=\rho_{{\bf k}}^{cv2}(t)$, the equations (\ref{Liouville}) for the matrix elements $\rho_{{\bf k}}^{cv1}(t)$ and $\rho_{{\bf k}}^{cv2}(t)$
in the third-order approximation (see Eqs.(\ref{VXC}) and (\ref{deltan2}))
correspond to the following equations for the polarization:
\begin{eqnarray}
&i&\frac{\partial}{\partial t}P_{1{\bf k}}(t)=\left[ 
\varepsilon_{{\bf k}}^{c}-\varepsilon_{{\bf k}}^{v}
\right]P_{1{\bf k}}(t)
\nonumber \\
&+&\sum_{{\bf q}}\int dt'\alpha_{11{\bf kq}}(t,t')P_{1{\bf q}}(t')
+{\bf d}_{{\bf k}}^{cv1}{\bf E}(t)
\nonumber \\
&+&\sum_{{\bf q,p,Q}}
P_{1q}^{*}(t)\int dt' F_{kqpQ}^{11}(t,t') P_{1p}(t')P_{1Q}(t')
\nonumber \\
&+&\sum_{{\bf q,p,Q}}
P_{2q}^{*}(t)\int dt' F_{kqpQ}^{12}(t,t') P_{1p}(t')P_{2Q}(t'),
\label{polarization}
\end{eqnarray}
and similar equation for the polarization $P_{2{\bf k}}(t)$ with interchange of $1$ and $2$.
In these equations,
\begin{eqnarray}
\alpha_{11{\bf kq}} (t,t')= 2 \int d{\bf r} \int
d {\bf r}' \varphi_{1{\bf k}}^{*}({\bf r})
f_{\rm xc}({\bf r},t;{\bf r}',t')
\varphi_{1{\bf k}}({\bf r}'),
\label{alpha}
\end{eqnarray}
and 
\begin{eqnarray}
F_{11{\bf kq}} (t,t')= \frac{1}{3!} \int d{\bf r} \int
d {\bf r}' \int d{\bf r}'' \int d {\bf r}''' 
\varphi_{1{\bf k}}^{*}({\bf r})\varphi_{1{\bf q}}^{*}({\bf r}')
\nonumber \\
\times
f_{\rm xc}''({\bf r},t;{\bf r}',t;{\bf r}'',t';{\bf r}''',t')
\varphi_{1{\bf p}}({\bf r}'')\varphi_{{\bf Q}}({\bf r}''')
\label{F11}
\end{eqnarray}
\begin{eqnarray}
F_{12{\bf kq}} (t,t')= \frac{1}{3!} \int d{\bf r} \int
d {\bf r}' \int d{\bf r}'' \int d {\bf r}''' 
\varphi_{1{\bf k}}^{*}({\bf r})\varphi_{2{\bf q}}^{*}({\bf r}')
\nonumber \\
\times
f_{\rm xc}''({\bf r},t;{\bf r}',t;{\bf r}'',t';{\bf r}''',t')
\nonumber \\
\times (\varphi_{1{\bf p}}({\bf r}'')\varphi_{{\bf Q}}({\bf r}''')
+\varphi_{1{\bf Q}}({\bf r}'')\varphi_{{\bf p}}({\bf r}'''))
\label{F12}
\end{eqnarray}
are the corresponding TDDFT electron-hole and two-electron/two-hole scattering (interaction) potentials.
In the last equations, $\varphi_{l{\bf k}}({\bf r})=\psi_{k}^{c(0)}({\bf r})\psi_{k}^{vl(0)}({\bf r})$
are the Kohn-Sham ''exciton" wave functions, and  $f_{\rm xc}''$ is the second derivative of the XC kernel
with respect to the charge density.
Similar equations can be written for the polarization $P_{2{\bf k}}(t)$ with interchange of $1$ and $2$
in Eqs.~(\ref{polarization})-(\ref{F12}).
Let us briefly discuss the meaning of the approximation which results in Eq.(\ref{polarization}).
This equation follows from the following nonequilibrium part of the XC energy:
\begin{eqnarray}
\delta E_{XC} &~&
\nonumber \\
&=&\frac{1}{2}\int d{\bf r}dt\int d{\bf r}'dt'\delta n({\bf r},t)f_{XC}({\bf r},t;{\bf r}',t')
\delta n({\bf r}',t')
\nonumber \\
&+&
\frac{1}{4!}\int d{\bf r}dt\int d{\bf r}'dt'\int d{\bf r}''\int d{\bf r}'''
\delta n({\bf r},t)\delta n({\bf r}',t)
\nonumber \\
&\times& f_{XC}''({\bf r},t;{\bf r}',t;{\bf r}'',t';{\bf r}''',t')
\delta n({\bf r}'',t')\delta n({\bf r}''',t').
\label{EXC}
\end{eqnarray}
Indeed, in this case one can express the fluctuation of the electron charge density
in terms of the polarizations $\delta n({\bf r},t)\simeq \sum_{l,{\bf k}} P_{l{\bf k}}(t)\varphi_{l{\bf k}}({\bf r})$
and substitute it into Eq.~(\ref{EXC}), which results in Eq.~(\ref{polarization}). 
In derivation of Eq.~(\ref{polarization}) we took into account only the terms
proportional to $P_{1}^{4}$, $P_{1}^{2}P_{2}^{2}$ and $P_{2}^{4}$ in the expression for the nonequilibrium
exciton thermodynamic potential, because these terms give the largest contribution to the energy due to the minimal number of oscillating factors $\exp (i\omega_{l{\bf k}}t)$ in front of each $P_{l{\bf k}}(t)$ ($\omega_{l{\bf k}}$ is the corresponding exciton frequency), in analogy to the case of the two-component
$\phi^{4}$ model \cite{phi4}. We assume that the time arguments in the first two and the last two 
densities in the last term in $\delta E_{XC}$ are equal. This approximation corresponds
to the mapping the effective action on the corresponding effective action of free biexcitons:
\begin{eqnarray}
\int d{\bf r}dt \int d{\bf r}'dt' B^{*}({\bf r},{\bf r}',t)D^{-1}({\bf r},{\bf r}',t;{\bf r}'',{\bf r}''',t')
\nonumber \\
\times B({\bf r}'',{\bf r}''',t),
\label{biexcitonmanybody}
\end{eqnarray}
assuming that  $B({\bf r},{\bf r}',t)\sim \delta n({\bf r},t)\delta n({\bf r}',t)$
and that the inverse biexciton propagator 
$D^{-1}({\bf r},{\bf r}',t;{\bf r}'',{\bf r}''',t')\sim f_{XC}''({\bf r},t;{\bf r}',t;{\bf r}'',t';{\bf r}''',t')$.

\section{The 2DFS and a possibility of the biexcitonic states}
\label{2DFS}

In order to find the 2D Fourier spectrum of the system in the case of a three-pulse excitation, one needs to find the third-order polarization $P_{l{\bf k}}^{(3)}$ given on the left-hand side of Eq.(\ref{polarization}). This can be done by inserting 
the solution for the linear polarizations into the right-hand side of the equation and then by making the Fourier transform with respect to the delay $\tau$ and the measurement $t$ times, which results in the spectrum that depends on the frequencies $\omega$ and $\Omega$, correspondingly. For simplicity, we will assume
that the initial pulse takes place at time $t_{1}=0$, while two other pulses occur at times $\tau >0$.
The first-order polarization can be found by solving the equation which consists of the first
two lines in  Eq.~(\ref{polarization}). The solution is
\begin{eqnarray}
P_{lk}(t)=\int_{-\infty}^{t}e^{-i(\omega_{lk}-i\Gamma_{lk})(t-t')}d_{k}^{cv1}E(t')dt',
\label{P1solution}
\end{eqnarray}
were $\omega_{lk}$ and $\omega_{lk}$ are the energy and the lifetime of the $l$-exciton.

In order to obtain biexcitons, one needs to consider a retarded exciton interaction $F(t-t')$ \cite{Ostreich3}.
Before presenting the numerical results of the application of our approach, it is useful to discuss the conditions which
$F(t-t')$ has to satisfy in order to obtain the biexcitonic peaks.
As it was argued in Ref.~\cite{Ostreich3}, for this purpose in the similar equation for the polarization in the many-body case, the memory function can be approximated by
\begin{eqnarray}
F(t)=\int_{0}^{\infty}d\omega \rho( \omega ) e^{-i\omega t},
\label{F2}
\end{eqnarray}
where "the heat bath" spectral density  $\rho (\omega )$ in the low-frequency limit, which defines the long-time asymptotic behavior of the system,
can be approximated by a power function $\rho (\omega )\sim \omega^{\alpha}$.
This function defines the dissipation processes for given exciton due to the environment consisting of the surrounding excitons. 
In the case, when $\alpha$ is smaller, equal or larger than $1$, the 
dissipation is called "sub-ohmic", "ohmic" and "super-ohmic", respectively \cite{Caldeira}. 
Since the spectral function must decay
at large frequencies, the general form of the spectral density was approximated by
\begin{eqnarray}
\rho (\omega )=A\omega^{\alpha}e^{-\omega/\omega_F},
\label{rho}
\end{eqnarray}
where $\omega_{F}$ is the frequency scale and A is the normalization constant.
In the case of a one-dimensional model for the excitons, the authors of Ref.~\cite{Ostreich3} found
$\alpha =1$. 

In TDDFT, the memory effects are described by the matrix elements
of the first and second derivatives of the XC kernel.
One can in principle construct a $f_{XC}''$ to reproduce
(\ref{F2}). In this case, the equation for the polarization and its solution, including the biexcitonic features,
 will coincide with the many-body case. While there are known constraints on the frequency-dependent
(non-adiabatic) part of $f_{XC}$ at small and large frequencies, the full dependence of  $f_{XC}''$ has not been discussed in detail so far.
In particular, it is known that the exact asymptotic of the XC kernel
at large frequencies is $f_{XC}\sim a+b\omega^{-2}$ \cite{vanLeeuwen}. In the low-frequency limit,
the information about the exact behavior of $f_{XC}$ is more limited. It is known that
it can have poles in the case of a finite system in the discrete part of the spectrum.
One can construct the frequency-dependence of the XC kernel for all ranges of frequencies by using results
of the homogeneous electron gas: 
$f_{XC}(\omega\rightarrow 0)\rightarrow 0, f_{XC}(\omega\rightarrow\infty)\rightarrow \omega^{-3/2}$
\cite{Marques}.
From these results one can suggest the following rather general form for the non-adiabatic part
of the XC kernel:
\begin{eqnarray}
f_{XC}(\omega)=A\frac{\omega^{\alpha}}{1+(\omega/\omega_{F})^{\alpha +\beta}},
\label{fxcHEG}
\end{eqnarray}
where $\alpha >0$  and $\beta =2$, though $\beta =3/2$ is also worth of special attention.
Since our main goal is to explore the role of non-adiabaticity in the nonlinear response, we assume that full kernel is the product of the frequency part Eq.(\ref{fxcHEG}) and the spatial part ${\tilde f}({\bf r},{\bf r}')$:
$f_{XC}({\bf r},\omega )=A[\omega^{\alpha}/(1+(\omega/\omega_{F})^{\alpha +\beta})]
{\tilde f}({\bf r},{\bf r}')$, 
which allows us to separate the spatial and temporal contributions to the interaction, and which makes the analysis more transparent.
More general case of the kernel may be considered, though we believe that main non-adiabatical effects
in the many-exciton system can be captured by this function.

It is also natural to assume that the retardation effects in the exciton case ($f_{XC}''$) are described by the same
time-dependence as in the electron case ($f_{XC}$), because the excitons are composed of "the elementary" quasi-particles, electrons and holes, interacting through the retarded interaction given above. Indeed, using the frequency-dependence
Eq.~(\ref{fxcHEG}) for $f_{XC}''$ one can demonstrate that the corresponding 2DFS includes biexcitonic features,
similar to the many-body approximation case Eqs. (\ref{F2}), (\ref{rho}) \cite{Ostreich3}.
However, using such an approximation gives rise to many questions about the physical meaning
of the corresponding spectral function, in particular: Can one assume that the heat-(or exciton-) bath dissipation
results in the formation of biexcitonic states? This question is nontrivial and deserve a deep study.
In this paper, we use a simplified form of $F(t)$ by making the analogy with the
many-body theory, where the biexciton propagator $D$ in Eq.~(\ref{biexcitonmanybody}) must contain poles at the biexciton frequencies.
This means that we postulate that such poles must be present in $f_{XC}''$.
Namely, we approximate  
\begin{eqnarray}
f_{XC}''({\bf r},t, {\bf r}',t',{\bf r}'',t'', {\bf r}''',t''')\simeq
g_{XC}(t-t'')
\nonumber \\
\times\delta (t-t')\delta (t''-t''')
{\tilde g}({\bf r},{\bf r}',{\bf r}'',{\bf r}''').
\label{g}
\end{eqnarray}
The approximation in the time-dependence of the kernel derivative
corresponds to taking into account only the two-particle interactions in the XC energy (\ref{EXC}).
Namely it is assumed that "the quasi-particle"
$\delta n({\bf r},t)\delta n({\bf r}',t)$ at time $t$ (two electrons at points ${\bf r}$, ${\bf r}'$) interacts with the one 
$\delta n({\bf r}'',t'')\delta n({\bf r}''',t'')$ at time $t''$ (two holes at points ${\bf r}''$, ${\bf r}''$),
which can be easily seen
from Eq.(\ref{EXC}):
\begin{eqnarray}
E_{XC}\sim\frac{1}{4!}\int d{\bf r}dt\int d{\bf r}'dt'\int d{\bf r}''\int d{\bf r}'''
\delta n({\bf r},t)\delta n({\bf r}',t)
\nonumber \\
\times
 g_{XC}(t-t''){\tilde g}({\bf r},{\bf r}',{\bf r}'',{\bf r}''')
\delta n({\bf r}'',t'')\delta n({\bf r}''',t'').
\label{EXC2}
\end{eqnarray}
In this case, the pole structure of $f_{XC}''(\omega )$ is contained in $g_{XC}(\omega)$: 
\begin{eqnarray}
g_{XC}(\omega )=
\sum_{l,k,m,q}\frac{A_{l,k,m,q}}{\omega-\omega_{Bl,k,m,q}+i\Gamma_{Bl,k,m,q}},
\label{Fpole} 
\end{eqnarray}
where $A_{l,k,m,q}$ is the weight, $\omega_{Bl,k,m,q}$+$\Gamma_{Bl,k,m,q}$ are the spectrum and the decay rate for the biexciton state
 formed by two excitons described by the orbital $l,m$ and the momentum ${\bf k},{\bf q}$ quantum numbers.
Indeed, the pole structure  (\ref{Fpole}) guarantees that the
largest contribution to $\delta E_{XC}$ (Eq.~(\ref{EXC2})) come from the states at the resonant frequencies,
e.g. the two excited-electron, two exciton, and in our case, the biexciton frequencies. 
This is physically reasonable situation, since we are assuming that there are only exciton or two-exciton (biexcitonic)
excitations in the system.
The imaginary part of the structure (\ref{Fpole}) 
maybe regarded as a particular case of the Lorentzian type kernel (\ref{fxcHEG}).

A pole structure of $f_{XC}$
capable to describe the multiple excitations with TDDFT was considered, for example, in Ref.~(\cite{24}). 
In this paper we propose a similar structure for the kernel second derivative $f_{XC}''$ Eq.(\ref{Fpole}). 
In Eq.(\ref{Fpole}) we again separate the spatial and temporal parts for the sake of simplicity of the analysis. 
From Eq.(\ref{Fpole})
one can easily the structure of the corresponding non-adiabatic part of the XC kernel:
\begin{eqnarray}
f_{XC}({\bf r}, t,{\bf r}',t')
\simeq\frac{1}{4!}\int d{\bf r}dt\int d{\bf r}'\int d{\bf r}''dt''\int d{\bf r}'''
\nonumber \\
\times\delta n({\bf r},t)\delta n({\bf r}',t)
g_{XC}(t-t''){\tilde g}({\bf r},{\bf r}',{\bf r}'',{\bf r}''')
\nonumber \\
\times\delta n({\bf r}'',t'')\delta n({\bf r}''',t'').~~~~~~~~~~~~~~~~~~~~~~~~~~~
\label{fXCpole2} 
\end{eqnarray}

\section{Solution}
\label{Solution}

We apply the formalism above to analyze the effect of the memory function on the two-dimensional
Fourier spectrum of a GaAs multiple quantum well and compare the results with experimental data \cite{Zhang}.
The result for the spectrum depends
on the the heavy and light exciton frequencies and lifetimes $\omega_{\mu}$ and $\Gamma_{\mu}$ ($\mu =h,l$) 
and the memory function $F(t)$. DFT calculations with the Quantum Espresso code \cite{23} with LDA potential
were used to generate the static wave functions (Since the width of the quantum well is usually
rather large ($>$10nm), we used the bulk approximation to generate the wave functions). With this input we used
also the experimental results for the  heavy and light exciton frequencies
 $\omega_{h}=1.539eV, \omega_{l}=1.546eV$ and for the heavy-light hole biexciton binding energy $E_{XX}=1.5meV$.\cite{Zhang} We use one value $\Gamma= 1meV$ for the exciton and biexciton lifetimes, which agrees with the experimental estimations.\cite{3}   Finally, we use $f_{XC}''$ defined (\ref{g}) defined by 
memory function (\ref{Fpole}) 
and for the the spatial part we us an adiabatic LDA approximation:
\begin{eqnarray}
{\tilde g}({\bf r},{\bf r}',{\bf r}'',{\bf r}''')
=\delta^{2}f_{X}^{LDA}({\bf r},{\bf r}')/\delta n({\bf r}'')\delta n({\bf r}''')
\nonumber \\
=f_{X}^{LDA''}({\bf r})
\delta ({\bf r}-{\bf r}')\delta ({\bf r}-{\bf r}'')\delta ({\bf r}-{\bf r}''').
\label{gtilde}
\end{eqnarray}
We compared the solutions with both adiabatic and non-adiabatic $g_{XC}$.
As the main result, it follows from our calculations that it is impossible to obtain the non-diagonal (biexciton) peaks using an instant (adiabatic) exciton-exciton interaction. 
In contrast, it is absolutely necessary to use the non-adiabatic exciton-exciton interaction in order to obtain two diagonal states (that correspond two heavy and two light excitons peaks) and the non-diagonal
(heavy-light exciton, as in co-circular excitation experiment\cite{Zhang}) bound state state in the 2DFS (Fig. 1). 
\begin{figure}[t]
\includegraphics[width=8.5cm]{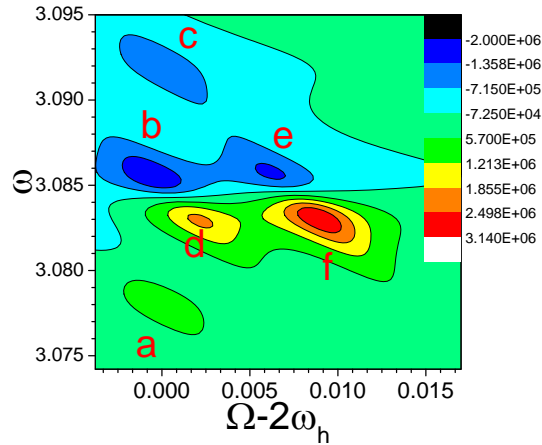}
\caption{\label{fig1} 
The 2D Fourier spectrum (the imaginary part) of the GaAs multi-well system. The Frequencies $\omega$ and $\Omega$ are given in meV. The features in the spectrum are marked as follows: a) two heavy excitons, b) a mixed (heavy+light) excitons, c) two light excitons; f) a mixed heavy-light biexciton; features d) and e) correspond to an exciton-biexciton coherence. This result corresponds to the kernel
(\ref{Fpole}) with the weights $A$ defined by static values (\ref{F11}) and (\ref{F12}) and the pole at the frequency $\omega_{Bl,k,m,q}=\omega_{h}+\omega_{l}-E_{XX}$
with the parameters given in the text (we neglected the momentum-dependence of the excitations).}
\end{figure}
Our results shown in Fig. 1 are in agreement with the experimental data \cite{Zhang}.

\section{Conclusions}
\label{Conclusions}

We have derived a TDDFT version of the nonlinear equation for the exciton dynamics.
We have
found that the effective time-dependent exciton-exciton interaction is defined by the non-adiabatic part of the XC kernel
and  discussed possible time-dependencies of this  interaction.
We applied the approach to study the 2DFS of a GaAs quantum well system
and showed that the main features of the spectrum, including biexcitons, can be reproduced within the TDDFT approach.
The density-matrix formulated version of TDDFT developed in the paper can be used to analyze the nonlinear ultrafast response of different types of many-fermion system. The technical simplicity and physical transparency of this ab initio approach makes it favorable comparing to the currently used many-body techniques.  In most cases, the non-adiabatic (memory) effects play a crucial role in the ultrafast response and the proposed theory when applied to
analyze the 2DFS and other ultrafast experiments may lead to an improvement
of our understanding of the structure of the non-adiabatic XC kernels for different types of systems
where these effects play an important role.

\section*{Acknowledgements}
We would like to thank Dr. Mikhail Erementchouk for numerous enlightening discussions.
This work was supported in part by grants
NSF-ECCS 072551, NSF-ECCS-0901784, NSF-ECCS-1128597, AFOSR No. FA9550-09-1-0450, DARPA/MTO
Young Faculty Award HR0011-08-1-0059  (M.N.L.)
and DOE-DE-FG02-07ER15842 (V.T.).

\end{document}